\documentstyle[11pt,aaspp4]{article}

\received{1997 June 6}
\accepted{1997 August 22}
\journalid{492}{1998 January 10}
\articleid{452}{460}

\slugcomment{To appear in The Astrophysical Journal, 492.}

\lefthead{Szokoly et al.}
\righthead{local galaxies}

\begin{document}

\title{The Near-Infrared Number Counts and Luminosity Functions\\
       of Local Galaxies }

\author{Gyula P. Szokoly}
\author{Mark U. Subbarao}
\author{Andrew J. Connolly}
\affil{Department of Physics and Astronomy, The Johns Hopkins University, 
    Baltimore, MD 21218, USA}
\and
\author{Bahram Mobasher}
\affil{Astrophysics Group,
    Blackett Laboratory,
    Imperial College,
    Prince Consort Road,
    London SW7 2BZ,
    U.K.
}



\begin{abstract}

This study presents a wide-field near-infrared (K-band) survey in two
fields; SA 68 and Lynx 2. The survey covers an area of 0.6 deg.$^2$,
complete to K=16.5. A total of 867 galaxies
are detected in this survey of which 175 have available redshifts. The
near-infrared number counts to K=16.5 mag. are estimated from the
complete photometric survey and are found to be in close agreement
with other available studies. The sample is corrected for
incompleteness in redshift space, using selection function in the form
of a Fermi-Dirac distribution. This is then used to estimate the local
near-infrared luminosity function of galaxies.  
A Schechter fit to the infrared data gives: 
M$^\ast_K = -25.1 \pm 0.3$, $\alpha = -1.3\pm 0.2$ and $\phi^\ast
=(1.5\pm 0.5)\times 10^{-3}$ Mpc$^{-3}$ (for H$_0=50$ Km/sec/Mpc and
q$_0=0.5$). When reduced to $\alpha=-1$, this agrees with other
available estimates of the local IRLF.
We find a steeper slope for the faint-end of the infrared 
luminosity function when compared to previous studies.
This is interpreted as due to the presence of a population of faint but 
evolved (metal rich)
galaxies in the local Universe. However, it is not from the same population
as the faint blue galaxies found in the optical surveys. 
The characteristic magnitude ($M^\ast_K$) of the local IRLF indicates that the
bright red galaxies ($M_K\sim -27$ mag.) have a space density of
$\le 5\times 10^{-5}$ Mpc$^{-3}$ and hence, are not likely to be local
objects.
 
\end{abstract}

\keywords{surveys -- cosmology: observations -- galaxies: photometry --
infrared: galaxies}

\section{Introduction}

The change in the intrinsic properties of galaxies as a function of
lookback time provides a sensitive probe of the theories of formation
of galaxies. A detailed knowledge of this is also required for
interpreting the deep counts of galaxies and hence, exploring the
underlying geometry of the Universe. At the optical, far-infrared and
radio wavelengths, however, the observed properties of galaxies are
sensitive to the morphological types, star formation history and dust
content, making it difficult to disentangle the effects of evolution
and cosmology at these passbands. The exception is the near-infrared
wavelength (2.2 $\mu m$) which is less affected by the evolutionary
processes (due to the dominance of the near-infrared light by old,
near-solar mass stars) and can be more securely modeled. Also, at
these wavelengths the galaxy spectral energy distributions are similar
for all galaxy types, leading to similar K-corrections for all
galaxies. Therefore, the near-infrared number-magnitude counts of
galaxies are less affected by the mix of galaxy types or uncertainties
due to evolution, compared to other wavelengths. However, the
interpretation of the infrared counts requires knowledge of the
faint-end slope of the local infrared luminosity function (IRLF).

There are other independent studies, also requiring knowledge of the
IRLF of galaxies.  Due to type dependence of the optical LF
(\cite{efs88}) and existence of a density-morphological type relation
for galaxies, the optical LF is expected to be
environmental-dependent. Indeed, recent spectroscopic studies of field
galaxies have shown the existence of a population of massive,
star-forming galaxies (\cite{cow95}; \cite{cow96}) in the field, not
present in clusters (\cite{tren97}). The IRLF is not sensitive to such
star-forming galaxies, which affect the bright-end of the optical LF,
and hence, is likely to have a universal form.  Further, this also
gives the integrated star formation in the local Universe.
 
Previous studies of the IRLFs of galaxies either depend on surveys
selected at optical wavelengths (\cite{mob93}) or, are mainly pencil
beam surveys, only covering small solid angles (\cite{cow96};
\cite{glaz94}). Also, the only study of the morphological type
dependence of the IRLF (\cite{mob93}) is biased against early-type
galaxies (due to the optical selection of galaxies) and has too few
objects of a given type. Moreover, since this sample is optically
selected, it is not possible to convert it to an infrared limited
survey because of the changes in type mix (i.e. color) with
redshift. Therefore, it can only constrain the bright-end of the IRLF.
Recently, a wide-angle near-infrared survey was performed
(\cite{gard97}) to a limit of K=15 magnitude. This is, however, too
shallow to constrain the faint-end slope of the local IRLF with the
statistics becoming very poor fainter than $M_K=-23.5 + 5 log
(H_0/50)$ mag (ie. one magnitude below $M^\ast$), where the
disagreement between different measurements of the IRLFs becomes the
largest.

To overcome these problems, we have carried out a medium deep
near-infrared survey, covering a large area and complete to $K\sim
16$ mag. The fields were selected to have optical multicolor data, with
redshifts available for a sub-sample of galaxies. In this first
paper of a series, we introduce this near-infrared selected survey. 
Using published redshifts, we then construct the local IRLF of
galaxies and compare it with other, independent measurements.  A study
of the morphological type-dependence of the IRLF, color-magnitude
relations for field galaxies and the bivariate optical/infrared LF
will be presented in following papers.

In section 2 we present the observations and data reduction.
Section 3 explains source detection and photometry. The astrometry and
construction of the infrared catalogue are discussed in section 4.
The number-magnitude counts are presented in section 5. Section 6 explores
the completeness of the sample. The infrared luminosity function is 
constructed in section 7. Finally, our conclusions are summarized in
section 8.

\section{Observations and Data Reduction}

The near-infrared observations presented here, were performed using
the IRIM camera on the 1.3m telescope at KPNO. This instrument is a
NICMOS 3 array (HgCdTe) with $256\times 256$ pixels. The pixel scale
of $1.96''$ corresponds to a field of view of $8.3'\times 8.3'$. This
configuration enables us to undertake a wide field survey with
relatively little cost in observation time.  The observations were
carried out in $K_s$ filter ($\lambda = 2.15 \ \mu m$; $\Delta\lambda =
0.33 \ \mu m$). This filter is {\em not} the K' filter described by
Wainscoat and Cowie (1992)\markcite{wain92} but the K-short filter also
used by the 2MASS survey -- unfortunately the IRIM instrument manual
refers to it as K'. The filter cuts off about $0.1\ \mu m$ bluer than the
standard K filter. This significantly reduces the thermal emission
resulting in a decrease of approximately 25\% in the sky background
over the standard K filter.

Two fields were selected for near-infrared observations, {\em Lynx-2}
($8^h 41^m 43.7^s, 44^\circ 46^\prime 42^"$, 1950) and {\em SA-68} ($0^h
14^m 53^s, 15^\circ 36^\prime 48^"$, 1950). The fields were selected to
have existing multicolor photometric ($UB_JR_FI_N$) and spectroscopic
data (\cite{munn97}). The near-infrared observations were carried
out by dithering around each pointing position. Using an offset of 20
arcsec and an exposure time of 1 minute (20 sec. with 3 co-adds), a
total of 5 exposures were taken, corresponding to an integration time
of 5 minutes per pointing. This scheme was chosen to avoid cosmic rays
and bad pixels. The dithered pointings were used to estimate the sky
background and to construct flatfields. This dithering pattern was
repeated a total of 36 times to produce a mosaic of each field. Small
overlaps between the mosaic elements were incorporated into the dither
pattern in order to check the internal consistency of the photometry
and astrometry.  Using an integration time of 5 minutes, we reach a
formal magnitude limit of $K_s=17$ for a 5$\sigma$ detection in an 5
arcsec aperture.  A total area of about 0.8 square degree was covered
in both fields to a depth of $K_s=17$.  The overlap between the optical
and near-infrared survey is about 0.6 square degree.

The data were reduced using standard IRAF routines. After
dark-subtraction, each exposure was flatfielded using the following
three steps: Initially the dome flat was used to correct the data
frames for the large-scale sensitivity variance of the detector.  The
flatfielded data frames were then corrected for changes in the
instrument response (which was found to vary slightly over a time
scale of a few minutes) by dividing them by the sky flats, constructed
by median filtering 5-10 successive data frames. Finally, the small
background gradients present on the images were estimated and removed
by creating a median filtered image with a very large box size and
subtracting this from the original data frames.

Bad pixels were removed by deriving the average signal and variance
for each pixel over the whole night and identifying those pixels that
were greater than (hot pixels) or less than (non-responding pixels) 5
$\sigma$ above or below the mean. Those pixels with a large variance
were identified as being due to variable sensitivity (i.e.\ noisy
pixels) and were excluded.

The final frames were registered to a common coordinate system using
bright stars, not affected by bad pixels, in the frame.  The measured
uncertainties of the centroiding of the bright stars was 0.1--0.2
pixels. Due to the undersampling of our data and the large number of
bad pixels, image shifts were done using linear interpolation after
excluding each of the identified bad pixels. This procedure
conserves the flux (which was our main goal) and maximizes the
signal-to-noise ratio. However, it does increase the intrinsic PSF of the
resultant images by about 40\%. The five exposures, covering the
same area, were then averaged and trimmed to construct the final
images.  Standard photometric calibrations were derived using the
UKIRT faint standards (\cite{cas92}).
The observations are estimated to have an intrinsic photometric
uncertainty of approximately 0.06 magnitude in the zeropoint.

\section{Source Detection and Photometry}

The source detection was performed using the SExtractor package
(\cite{sex}). Objects were identified using a Gaussian detection
kernel with FWHM of $3''$. A detection limit was set such that the
objects have a minimum of 5 pixels $1.5\sigma$ above the level of the
the sky background (after convolution). This corresponds to a surface
brightness of 19.7 mag arcsec$^{-2}$. For all detected objects, two
sets of magnitudes were measured, an aperture magnitude over a radius
of $5''$ and a Kron magnitude with a Kron parameter of 2.5
(\cite{kron80}). Due to the large pixel scale and the relatively small
intrinsic sizes of the galaxies at $K_s = 15$, the difference between
the aperture and Kron magnitudes were found to be negligible for the
majority of our sample.  At the brightest magnitude limit considered
in our survey, $K_s =13$, the galaxy sizes exceed the aperture size and
hence, are better represented by the Kron magnitudes. For the
subsequent analysis in this study, we use the Kron magnitudes. For the
adopted Kron parameter, these roughly correspond to the total
magnitudes of galaxies. No shape parameters (e.g.\ second moments)
were estimated from the near-infrared images because of the poor
spatial resolution and the existence of higher resolution optical
data.

  From the selection function derived for the near-infrared data we
estimate that our 5$\sigma$ detection limit is 17.0 magnitude and our
completeness limit is $K_s=16.5$. To determine the internal accuracy of
our photometry, those sources detected in the overlapping regions are
compared in Fig. 1. The rms dispersion of the one-to-one correlation is 0.15
magnitude at the completeness limit and 0.3 magnitude at the detection
limit. These are taken as the internal uncertainties in our infrared
photometry. The external photometric accuracy is estimated by
comparing, in Fig. 1, the $K_s$ band magnitudes for those galaxies in
common between the present study and those by Bershady et al. (1994). This
gives a larger rms scatter of 0.2 mag which is, in part, due to the
fact that we are comparing our Kron magnitudes with the aperture
magnitudes measured by Bershady (with associated aperture
corrections). The spatial resolution of our data prevents an analysis
of the Bershady data using matched apertures. It is encouraging,
however, that we detect no significant zero point offset or
correlation that could suggest systematic errors in either study.  For
the rest of this paper we assume an internal photometric accuracy of
0.15 mag.

\section{Astrometry and Comparison with the Optical Catalogue}

As noted above, the large pixel scale of the near-infrared detector here, 
limits our ability to perform accurate star galaxy separation (essentially
all but the largest objects detected are point-like). However, 
for each of the survey fields, deep independent multicolor surveys exist, 
which can be used to classify the near-infrared detections (\cite{koo86};
\cite{kron80}).  This requires accurate astrometric calibration of our
detected sources. 

The astrometry was carried out in two steps.  Initially, approximate
astrometric offsets were determined by identifying stars on each frame
from the Guide Star Catalog (GSC; \cite{gsc1}; \cite{gsc2}; \cite{gsc3}).
This resulted in 2-6 stars
per frame. Averaging over all the frames, we found the overall
deviation from a zero-distortion focal plane negligible. An astrometric
zero-point for each frame was determined from the GSC stars. Using
this crude calibration, the near-infrared data were paired up with the
optical catalogs by searching within a $10''$ radius around each
detected object. From the paired data, we extract isolated sources
(i.e.\ those near-infrared objects with only one optical counterpart
within the search radius) and rederive a more accurate astrometric
solution. While the astrometric solution for each frame can be
established rather accurately ($0.1''-0.2''$), the uncertainty in
astrometry for individual objects remains large ($\sim 1.8''$), due to
the large PSF.

We construct an optical--near-infrared catalogue of all the detected
objects in the overlapping region, by extending the technique of
Sutherland and Saunders (1992)\markcite{suth92} to use both positional
and magnitude information. The simplest method to identify the same
object in different catalogs is to estimate the maximum positional
uncertainty and pair-up the detected objects with separations smaller
than this threshold. This technique works well for surveys where the
average astrometric uncertainty is much smaller than the typical
separation between the objects.  However, for moderately dense fields,
this method does not produce reliable catalogs as shown for our
dataset in Figure 3a. The dashed line is the number of near-infrared
detections without an optical counterpart and indicates that a
tolerance level of $\sim 2$ normalized distance units is required
(normalized distance is $\sqrt{(d_1/\sigma_1)^2+(d_2/\sigma_2)^2}$,
where $d_1$ and $d_2$ are the angular positional differences along the
two axes with $\sigma_1$ and $\sigma_2$ their mean respective
uncertainties). At this level, about 25\% of the detections have
multiple optical counterparts (dotted line). The solid line shows the
number of cases where this simple method results in unique
matches. This cuts-off rather sharply, indicating that this
identification is not too reliable.

To eliminate multiple hits within the matched catalogs the usual
practice is to pick the closest object from the optical catalog. One
can assign a {\em likelihood} to this identification
\begin{equation}
L={q(m)f(x,y)\over n(m)}
\end{equation}
where $q(m)$ is the probability distribution in magnitude, $f(x,y)$
describes the distribution of positional uncertainties and $n(m)$ is the
surface density of objects of magnitude $m$. This method does not take
into account the local information. For isolated objects, the match is
likely to be genuine even if the positional uncertainty is large, while in a
crowded area multiple objects can be very close, making the
identification unreliable (even though the likelihood is high). Following 
Sutherland and Saunders (1992)\markcite{suth92}, we address this problem by
defining a new local quantity, the {\em reliability}, as
\begin{equation}
R={L_i\over\sum\limits_j L_j + (1-Q)}
\end{equation}
where $L_i$ is the likelihood of an identification, as defined above,
$L_j$ are the likelihoods of other candidates and $Q$ is the
probability that the identification is possible (i.e. both catalogs
contain the object). This formalism properly takes into account the
existence of other identified candidates.

In the present study, we extend this technique to include positional
and magnitude information. First, we identify isolated objects
(sources with no other detection within a $10''$radius) present in
both the catalogs. Since the infrared survey is not as deep as its
optical counterpart, this subset, in practice, contains only genuine
identifications. From these data we then derive a correlation between
the near-infrared ($K_s$-band) and optical detections by fitting the
function $K^\ast (U,B_J,R_F,I_N)$ to a second degree polynomial
(\cite{conn95}). To allow for differences in optical photometry
between the two fields, this procedure was done independently for each
field.  The correlation between the predicted and measured K-band
magnitudes is shown in Fig. 2. The mean residual uncertainty for both fields
is 0.45 mag. and the error distribution is Gaussian with no outlying
points. Since about 40\% of the galaxies are isolated, we do not expect more
than a few (at most five) galaxies with unusual K-magnitude in the full
survey.

Using the fit $K^\ast (U,B_J,R_F,I_N)$, a normalized distance metric
is then defined as
\begin{equation}
\rho =\sqrt{{(\alpha_K-\alpha_0)^2\over\sigma_\alpha^2} +
{(\delta_K-\delta_0)^2\over\sigma_\delta^2} +
{(K-K^\ast)^2\over\sigma_K^2}}
\end{equation}
where $\sigma_\alpha, \sigma_\delta, \sigma_K$ are the uncertainties in
the spatial and color dimensions.
Since the distributions in all three directions are Gaussian and independent,
the likelihood functions are then calculated as
\begin{equation}
L = {f(\alpha,\delta,K)\over n(K)} \approx Q(< K) {e^{-\rho^2/2}\over 
2\pi \sigma_\alpha\sigma_\delta\sigma_K}
\end{equation}
where $n(K)$ is the density of optical detections and $Q(<K)$ the 
probability that an object of magnitude K has an optical counterpart
(in this case $Q(<K)\approx 1$ for the infrared catalogue). Finally, the
reliability of the match is defined as 
\begin{equation}
R_j = {L_j\over \Sigma_i L_i + (1-Q)}\approx {L_j\over \Sigma_i L_i}
\end{equation}
This technique is used to cross-compare the near-infrared and optical 
catalogs and as demonstrated in Figure 3b, it considerably improves the
match between the two surveys. 
The ratio of number of objects uniquely matched to the number of
objects with multiple matches is, at all distances, greater than that
derived when using only the positional information. Therefore,
including color information greatly reduces the number of spurious pairs, 
resulting in a much more reliable catalog.

For all the matched objects, the most reliable optical detection was
entered into the final catalogue. The calculated reliabilities, in
most cases, were found to be close to 1. For each object with an
optical counterpart, we adopt the classification (star or galaxy) as
defined in the optical catalog. An area of 0.8 deg.$^2$ was surveyed
in the near-infrared, of which 0.6 deg.$^2$ overlaps with the
multicolor optical surveys. Over this area, a total of 871 objects
were detected in the near-infrared survey, of which 867 have optically
identified counterparts. The 4 objects without an optical counterpart
are likely to correspond to faint spurious detections.

The final infrared catalogue contains 867 galaxies to $K_s=17$ mag.
All these objects have optical data. The redshifts are compiled from
literature for a sub-set of galaxies in the infrared catalogue, giving
a total of 175 galaxies with measured redshifts (\cite{munn97}).
This sample will be used in the following sections to construct
the number counts and near-infrared luminosity function of local
galaxies.

Using the estimated K-band magnitudes, we verify that our survey is
shallow enough such that blending of close pairs of galaxies does not
significantly affect the catalog. We calculate the expected infrared
magnitude for every optical galaxy around each infrared detection. We
individually examine all infrared detections that have multiple
optical counterparts within 10 arcsec and with $K_{predicted} <16.5$
(i.e.\ sufficiently bright that they would make it into the K selected
sample). Within the complete photometric survey there were 10 such
candidates each of which was handled correctly by the photometry
package.

\section{The Near-infrared Number Counts}

The near-infrared number counts, estimated from this study, are
presented in Table 1. These are compared in Fig. 4 and Table 2 with
the counts from other independent studies as compiled by
Gardnet et al. (1993). The comparison between the various studies is made
over the magnitude range covered in this study ($K_s=13-16.5$ mag.). In
all the counts in Figure 4, the uncertainties are estimated using the
technique developed by Gehrels (1986)\markcite{gehr86}.

The counts derived from our two independent fields, SA68 and Lynx2,
agree within their uncertainties over the entire magnitude range covered
here. They are also in good agreement with the counts in Gardner (1996)
\markcite{gard96} and Huang and Cowie (1997)\markcite{huang97b}, each covering
a larger area (9.8 deg$^2$) but to a shallower depth ($K<16$). These results
confirm that the present near-ir survey is complete to $K_s=16.5$
mag. Moreover, it indicates that there is no anomaly, due to density
enhancement in our fields or photometric zero-points, affecting the
near-ir distribution of galaxies in this study.

Fitting a linear relation to the number-magnitude counts for $14.5 <
K_s < 16.5$ we derive a slope of $0.46 \pm 0.06$, $0.53 \pm 0.04$ and
$0.50 \pm 0.03$ for the SA68, Lynx2 and combined samples respectively.
These values are lower than those derived by Huang and Cowie (1997)
\markcite{huang97b} and Gardner (1997)\markcite{gard97} from their wide field
survey. They are also lower than the predictions by Huang et al. (1997)
\markcite{huang97} for a no evolutionary
model.  The shallower slope may be indicative of an overdensity of low
redshift galaxies in our two fields (which is clearly seen in the
redshift distribution of the Lynx2 field but not in SA68). It is more
likely, however, that the variation amongst the measured slopes is an
artifact of fitting linear relations over a narrow range of
magnitudes.

\section{Selection Functions}

As the spectroscopic observations in this
study are based on an optically selected sample, it is likely that at
fainter $K_s$ magnitudes, the survey becomes incomplete in redshift
space. Therefore, it is instructive to estimate the incompleteness in
redshift as a function of $K_s$. Assuming a Fermi-Dirac distribution
for the selection function (\cite{STY}),
$\left(\exp\left({m-m_l\over\Delta m}\right)+1\right)^{-1}$, the
incompleteness (defined as the change with magnitude of the ratio of
the number of galaxies with measured redshift to the total number of
galaxies to a given $K_s$ limit), is calculated and fitted to this
parametric form in Fig. 5. We find $m_l=15.25$, $\Delta m_l = 0.80$
for SA68 and $m_l=14.86$, $\Delta m_l = 0.54$ for Lynx2. For the
luminosity function analysis in this study, we select a redshift
completeness limit of $30\%$ in Fig. 5. This corresponds to a limiting
magnitude of $K_s=16$ and $K_s=15.25$ in the SA68 and Lynx2 fields
respectively. These will be used as the completeness limits in the
following section to construct the near-ir luminosity function. The
redshift distributions for the complete sample, containing both the
fields, are presented in Fig. 6. There are a total of 110 galaxies
in the two fields brighter than the above $K_s$-band magnitude limits
in this survey. Assuming the estimated magnitude
limits, a $V/V_m$ test (\cite{Vmax}) gives $<V/V_m> =0.50 \pm 0.02$
and $<V/V_m> =0.35 \pm 0.02$ for the SA68 and Lynx2 fields
respectively. The smaller than expected $<V/V_m>$ for the Lynx2 field is
likely to be due to a cluster at $z\sim 0.05$ in this field
(Fig. 6). The effect of this density enhancement on the IRLF will be
explored in the next section. The redshift distribution in Figure 6
also shows that the galaxies in the infrared survey here are mainly
local ($z < 0.4$) objects.

\section{Local Near-ir Luminosity Function}

Most determinations of the luminosity function assume a parametric
form (normally a Schechter form) to fit to the observed data. The
shape of the luminosity function is, however, likely to be
type-dependent or be affected by density enhancements or its local
environment.  For these reasons, and because of the presence of a
cluster at $z=0.05$ in the Lynx2 field, we first use the
non-parametric C-method (\cite{Cmeth}) to assess the shape of the
local near-ir luminosity function. A parametric maximum likelihood
method (\cite{STY}), insensitive to density enhancement, will then be
used to find the best fit to the data.

Using the near-infrared K-corrections from Glazebrook et al. (1995)
\markcite{glaz95}, the K-band
luminosity function, estimated from the C-method is shown in Fig. 7.
The uncertainties are estimated using bootstrap re-sampling simulations.
Clearly, the shape of the local infrared luminosity function here is
consistent with a Schechter form (\cite{sch}). A parametric fit to
the data, using the cluster-free maximum likelihood method, gives
$M^\ast_{K_s} = -25.05 + 5 log (H_0/50)$ and $\alpha = -1.27$.  The
correlated uncertainty contours at $68\%$, $90\%$ and $95\%$ levels are
estimated for $M^\ast_{K_s}$ and $\alpha$ and presented in Fig. 8.
From these, we estimate 3$\sigma$ uncertainties corresponding to 0.3 mag. and
0.2 in $M^\ast_{K_s}$ and $\alpha$ respectively.  

To explore the sensitivity of this result to the density enhancement
in the Lynx2 field, we also find the luminosity function using the
`conventional' technique. This calculates the contribution 
from each galaxy to the volume covered by that galaxy at the apparent
magnitude limit of the survey (1/V$_{max}$). The total contribution of
galaxies in absolute magnitude intervals ($\sum_{i=1}^n\ 1/V_{max}^{(i)}$)
is then estimated and compared with the result from the cluster-free 
C-method in Figure 7. A Schechter function fit to these data, also
shown in Figure 7, gives: M$^\ast_{K_s} = -24.90$ mag., $\alpha =-1.42$ and 
$\phi^\ast = 9.5\times 10^{-4}$ Mpc$^{-3}$. 
Within the uncertainties, this is similar to the parameters estimated from the 
cluster-free maximum likelihood method, indicating that the presence of
non-homogeneities in our sample does not
significantly affect the resulting luminosity function. Therefore, 
the parameters
from the C-method/cluster-free fit here will be taken as the values for the
local near-infrared luminosity function in this study.

The normalization of the luminosity function ($\phi^\ast$ in the
Schechter formalism), not given by the cluster-free maximum likelihood
technique, is estimated using three different methods. First, performing a 
$\chi^2$ minimization of the Schechter function to the results from the
C-method in Figure 7. Second, using the $\phi^\ast$ value estimated
from the `conventional' technique, as presented above. 
Third, employing our estimate of $M^\ast_{K_s}$ and $\alpha$ to
establish a model of the K-band number counts and normalizing this
to the observed counts in Fig. 4. We adopt the normalization given by
the C-method and estimate the uncertainty based on the variation in $\phi^\ast$
derived by the three methods:
$\phi^\ast = (0.15\pm 0.05)\times 10^{-2} (H_0/50)^3$ Mpc$^{-3}$.

The IRLFs in Figure 7 are compared with a similar study by Gardner et al
(1997). Their measured IRLF is based on a larger sample ($\sim 500$ galaxies)
but is less deep than the present survey (it is only complete to $K\sim 15$
mag.). Although both the bright-end and normalizations agree fairly closely, 
the faint-end of the IRLF in this study is significantly steeper. 
The reality of the steep faint-end slope for the local IRLF, found here, 
will be further investigated by measuring redshifts for the fainter galaxies
in our sample, extending the completeness of our survey to $K_s\sim 17$ mag. 

In order to avoid the
correlation between $M^\ast$ and $\alpha$ affecting the fit and to
compare the present luminosity function with other independent studies, 
we fix $\alpha=-1$ and estimate $M^\ast_{K_s}$. The local IRLFs, estimated
by different groups, are presented in Table 3. The characteristic magnitude
($M^\ast_{K_s}$) here, is transformed from $K_s$ to K, using the relation in
section 5. Following Glazebrook et al. (1995)\markcite{glaz95}, we
apply a correction of $+0.22$ mag. to Mobasher et al. (1993)
\markcite{mob93} measurement to account for differences in
K-corrections. Also, an aperture correction of $-0.3$ magnitude  is applied to
\cite{glaz95} to convert it to the same scale as other
measurements. All the estimates are corrected to $H_0=50$ Km/sec/Mpc. 
It is clear from Table 3 that, at a given $\alpha$, the $M^\ast_K$ values
from different methods are in close agreement. The space density
of local galaxies ($\phi^\ast$) in this study is slightly smaller
than other similar measurements ($\sim 1.5$). This will be further explored by 
increasing the size of our sample, extending its 
completeness in redshift space. 

The steep faint-end slope of the local IRLF, if confirmed, will have
important implications towards constraining the models for formation
of nearby, low-luminosity field galaxies (ie. the mergers scenarios)
since the IRLF is mainly sensitive to the mass function and not the 
star formation (young population) in galaxies. Also, this implies the
existence of a large population
of evolved, metal rich  galaxies in the local Universe. 
The color distribution of the K-selected surveys shows that
the faint blue galaxies start to contribute to the galaxy counts at
about $K\sim 18$ mag. (\cite{gard95}). Such near-infrared 
surveys will then
reveal if the faint blue galaxies have an underlying population of old
stars. The surveys selected in the K-band are mainly dominated by
normal massive galaxies. Therefore, the characteristic magnitude
($M^\ast_K$) found for the IRLF here implies that the very red
galaxies with $M_K\sim -27$ mag. (\cite{ega96}; \cite{gra96})
have a space density of $\le 5\times 10^{-5}$ Mpc$^{-3}$ and
hence, are not likely to be local objects ($z < 0.4$).

We are currently completing the redshift measurements for the fainter
galaxies ($K_s > 16$) in the present survey. This will be used to
further constrain the faint-end slope of the `local' infrared
luminosity function by improving the statistical significance of the
sample and to explore its morphological type dependence.
 
\section{Conclusions}

In this study we carried out a wide-angle near-infrared galaxy survey in
two fields, SA68 and Lynx2. The survey is complete to $K_s=16$ (SA68)
and $K_s=15.25$ (Lynx2) and covers a total area of 0.6 deg.$^2$.
Matching the near-infrared detections with existing optical multicolor
surveys we derive a catalog of 867 galaxies. Of these, 175 have
available redshifts. Employing the complete photometric survey, the
near-infrared number-magnitude counts are estimated to $K\sim 16.5$
mag. and found to be in agreement with other independent measurements.

Correcting the spectroscopic samples for incompleteness in redshift
space, using selection functions in the form of Fermi-Dirac
distribution, the luminosity function has been derived for local
galaxies. Applying a parametric Schechter fit we find
$M^\ast_K = -25.09 \pm 0.3$, $\alpha = -1.27\pm 0.2$ and $\phi^\ast
=(0.15\pm 0.05)\times 10^{-2}$ Mpc$^{-3}$ (for H$_0=50$ Km/sec/Mpc and
q$_0=0.5$). When reduced to $\alpha=-1$, this agrees with other
available estimates of the local IRLF.

The most important limitation of our survey is that the redshift
survey is optically selected. Indeed we see a systematic, color
dependent incompleteness at $K_s>14.5$ in the Lynx-2 field and
$K_s>15$ in the SA-68. To estimate the effect of this
approximation, we used the simulated catalogs produced by Gronwall and
Koo (1995). We calculate the K-band luminosity function for an ideal,
100 percent complete, K-selected catalog and for an optically selected
redshift catalog with a selection function somewhat more conservative
than that of the KPGRS data ($R<18.5$).  We found that the two
calculated luminosity functions are identical, within the
uncertainties, up to $M_{K_s}=-22.5$.  For $M_{K_s}>-21.5$
the optically selected sample systematically overestimates the
amplitude of the luminosity function. We, therefore, limit our
analysis to $M_{K_s}=-22.5$ for the current data set.

The faint-end slope of the IRLF here is steeper than previous studies. 
This implies the presence of a population of faint but evolved (metal rich)
galaxies in the local Universe. Comparison with the flat faint-end slope of
the optical luminosity function reveals that this is unlikely to be the same
population as the faint blue galaxies found in the optical surveys. 
The characteristic magnitude ($M^\ast_K$) of the local IRLF indicates that the
bright red galaxies ($M_K\sim -27$ mag.) have a space density of
$\le 5\times 10^{-5}$ Mpc$^{-3}$ and hence, are not likely to be local
objects.

\acknowledgments 

We thank Mark Dickinson and Alex Szalay for useful
discussions about the analysis and interpretation of the near-infrared
data. We are grateful to David Koo, Richard Kron, Jeffrey Munn, Steven Majewski,
Matthew Bershady, and John Smetanka for pre-publication access to the
KPGRS catalogs.


\begin{deluxetable}{lrrrrrrrr}
\tablecaption{$K$-Band Differential Number Counts\label{tbl-1}}
\tablehead{
\colhead{Survey} & \colhead{K}   & \colhead{Raw $N$}   & \colhead{Low} & 
\colhead{High}  & \colhead{Area (arcmin$^2$)} & 
\colhead{N/mag/deg$^2$} 
} 
\startdata
SA-68 field      & 13.0 &   1 & 0.17  & 3.3 &  667  & 10.8  \nl
                     & 13.5 &   1 & 0.17  & 3.3 &       & 10.8  \nl
                     & 14.0 &   2 & 0.71  & 4.6 &       & 21.6  \nl
                     & 14.5 &  11 &  7.7  &  15 &       &  119  \nl
                     & 15.0 &  26 &   21  &  32 &       &  281  \nl
                     & 15.5 &  39 &   33  &  46 &       &  421  \nl
                     & 16.0 &  69 &   61  &  78 &       &  745  \nl
                     & 16.5 & 120 &  109  & 132 &       & 1295  \nl
Lynx-2 field     & 10.5 &   1 & 0.17  & 3.3 & 1519  &  4.7  \nl
                     & 11.0 &   0 &    0  & 1.8 &       &    0  \nl
                     & 11.5 &   1 & 0.17  & 3.3 &       &  4.7  \nl
                     & 12.0 &   5 &  2.8  & 8.4 &       &   24  \nl
                     & 12.5 &   1 & 0.17  & 3.3 &       &  4.7  \nl
                     & 13.0 &   6 &  3.6  & 9.6 &       &   28  \nl
                     & 13.5 &   6 &  3.6  & 9.6 &       &   28  \nl
                     & 14.0 &   7 &  4.4  &  11 &       &   33  \nl
                     & 14.5 &  26 &   21  &  32 &       &  123  \nl
                     & 15.0 &  42 &   36  &  50 &       &  199  \nl
                     & 15.5 &  81 &   72  &  91 &       &  384  \nl
                     & 16.0 & 160 &  147  & 174 &       &  758  \nl
                     & 16.5 & 270 &  254  & 287 &       & 1279  \nl
Combined fields      & 13.0 &   7 &  4.4  &  11 & 2185  &   23  \nl
                     & 13.5 &   7 &  4.4  &  11 &       &   23  \nl
                     & 14.0 &   9 &  6.1  &  13 &       &   30  \nl
                     & 14.5 &  37 &   31  &  44 &       &  122  \nl
                     & 15.0 &  68 &   60  &  77 &       &  224  \nl
                     & 15.5 & 120 &  109  & 132 &       &  395  \nl
                     & 16.0 & 229 &  214  & 245 &       &  755  \nl
                     & 16.5 & 390 &  370  & 410 &       & 1285  \nl
\enddata

 
\tablenotetext{}{
Listed are the field, the $K$-magnitude
(center of the bin), the raw number of galaxies, the upper and lower
1 $\sigma$ limits for the raw counts (\cite{gehr86}), the survey area 
and the number per magnitude per square degree.
}

\end{deluxetable}
\begin{deluxetable}{lrrrrrrrr}
\tablecaption{$K$-Band Differential Number Counts\label{tbl-2}}
\tablehead{
\colhead{Survey} & \colhead{K}   & \colhead{Raw $N$}   & \colhead{Low} & 
\colhead{High}  & \colhead{Area (arcmin$^2$)} & 
\colhead{N/mag/deg$^2$} 
} 
\startdata
Gardner '93 (HWS)   & 12.5 &   5 &  2.8  & 8.4 & 5690  &0.0009 \nl
                     & 13.5 &  23 &   18  &  29 &       &0.004  \nl
                     & 14.5 & 124 &  113  & 136 &       &0.02   \nl
Gardner '93 (HMWS) & 12.75 &   2 & 0.71  & 4.6 & 582.03& 24.7  \nl
                    & 13.25 &   0 &    0  & 1.8 &       &    0  \nl
                    & 13.75 &   1 & 0.17  & 3.3 &       & 12.4  \nl
                    & 14.25 &   3 &  1.4  & 5.9 &       & 37.1  \nl
                    & 14.75 &  11 &  7.7  &  15 &       &  136  \nl
                    & 15.25 &  22 &   17  &  28 &       &  272  \nl
                    & 15.75 &  45 &   38  &  53 &       &  557  \nl
                    & 16.25 &  89 &   80  &  99 &       & 1100  \nl
                    & 16.75 & 158 &  145  & 172 &       & 1900  \nl
Gardner '93 (HMDS) & 13.75 &   2 & 0.71  & 4.6 & 167.68&   86  \nl
                    & 14.25 &   0 &    0  & 1.8 &       &    0  \nl
                    & 14.75 &   1 & 0.17  & 3.3 &       &   43  \nl
                    & 15.25 &   5 &  2.8  & 8.4 &       &  215  \nl
                    & 15.75 &  21 &   16  &  27 &       &  902  \nl
                    & 16.25 &  28 &   23  &  34 &       & 1200  \nl
                    & 16.75 &  48 &   41  &  56 &       & 2060  \nl
Glazebrook '93       & 13.5 &   3 &  1.4  & 5.9 & 551.9 & 0.005 \nl
                     & 14.5 &   8 &  5.2  &  12 & 551.9 & 0.0145\nl
                     & 15.5 &  62 &   54  &  71 & 551.9 & 0.112 \nl
                     & 16.5 & 168 &  155  & 182 & 551.9 & 0.304 \nl
Gardner '96        & 10.25 & 1   &  0.17 & 3.3 & 30750 &$6\times10^{-5}$\nl
                    & 10.75 & 1   &  0.17 & 3.3 &       &$6\times10^{-5}$\nl
                    & 11.25 & 4   &   1.4 & 7.1 &       &$2.6\times10^{-4}$\nl
                    & 11.75 & 13  &   9.4 &  18 &       &$8.4\times10^{-4}$\nl
                    & 12.25 & 22  &    17 &  28 &       & 0.0014\nl
                    & 12.75 & 33  &    27 &  40 &       & 0.0021\nl
                    & 13.25 & 66  &    58 &  75 &       & 0.0043\nl
                    & 13.75 & 138 &   126 & 151 &       & 0.0090\nl
                    & 14.25 & 273 &   256 & 290 &       & 0.0177\nl
                    & 14.75 & 642 &   617 & 668 &       & 0.0418\nl
                    & 15.25 & 1290&  1250 &1330 &       & 0.0839\nl
                    & 15.75 & 2609&  2560 &2660 &       & 0.1697\nl
\enddata

 
\tablenotetext{}{
Listed are the survey, the $K$-magnitude
(center of the bin), the raw number of galaxies, the upper and lower
1 $\sigma$ limits for the raw counts, the survey area 
and the number per magnitude per square degree.
}

\end{deluxetable}

\clearpage
\begin{deluxetable}{lrrrr}
\footnotesize
\tablecaption{Schechter parameters for near-infrared luminosity function}
\tablewidth{0pt}
\tablehead{
\colhead{study} &
\colhead{$M^\ast_K$} & \colhead{$\alpha$} & \colhead{$\Phi^\ast$} &\colhead{n}}
\startdata

This study (C-method) & $-25.09$ & $-1.27$ & $0.15\times 10^{-2}$ & 110\\
This study (conventional method) & $-24.94$ & $-1.42$ & $0.95\times 10^{-3}$ &110\\
This study $\alpha = -1$ & $-24.84$ & $-1$     & $0.15\times 10^{-2}$ & 110\\
\cite{gard97}  & $-24.87$ & $-1.03$ & $0.22\times 10^{-2}$ & 532\\
\cite{glaz95}  & $-24.55$ & $-1.04$     & $0.33\times 10^{-2}$ & 98\\ 
\cite{mob93}   & $-24.88$ & $-1$     & $0.14\times 10^{-2}$ & 95\\  
\enddata

 
\tablenotetext{}{
}
\end{deluxetable}

\clearpage

\begin{figure}
\plotone{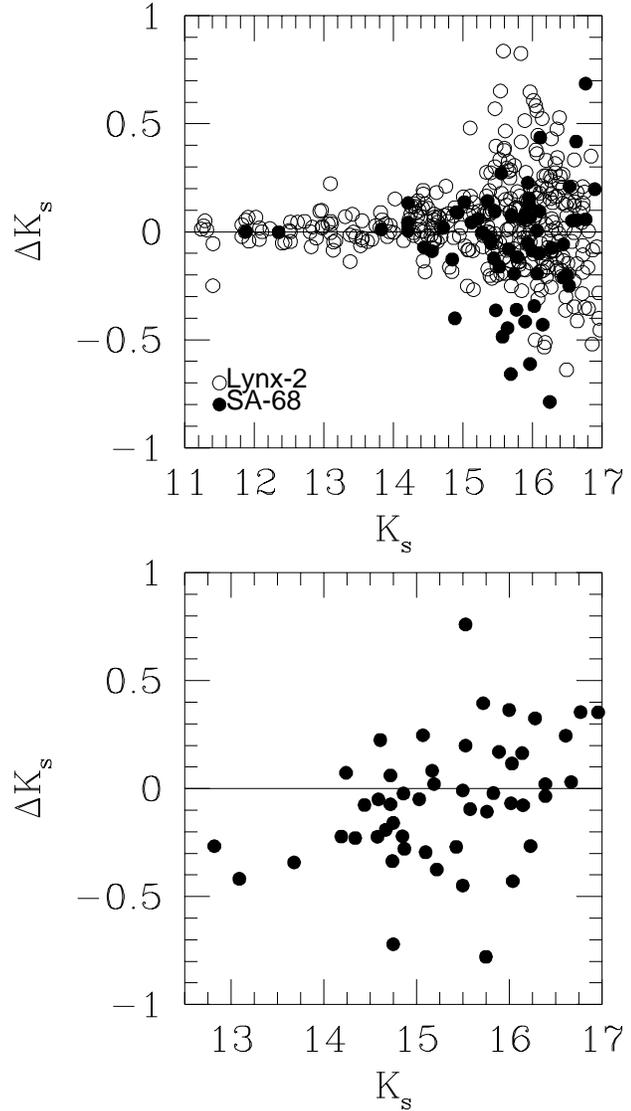}
\caption[internal.eps]{The photometric uncertainties within our data
are estimated internally from a comparison between repeated
observations and externally using published K-band photometry.  The
top panel shows the dispersion about a one-to-one correlation for
those sources detected in overlap IRIM frames. At the completeness
limit ($K_s = 16.5$) the rms dispersion is 0.15 magnitude. The lower panel
shows a comparison between our photometry and that of Bershady
(\cite{bers94}), $K_s^{(1)}$.  There is no significant offset between
our photometric system but due to the use of different magnitudes the
rms dispersion increases to 0.2 magnitudes. \label{fig1}}
\end{figure}

\begin{figure}
\plotone{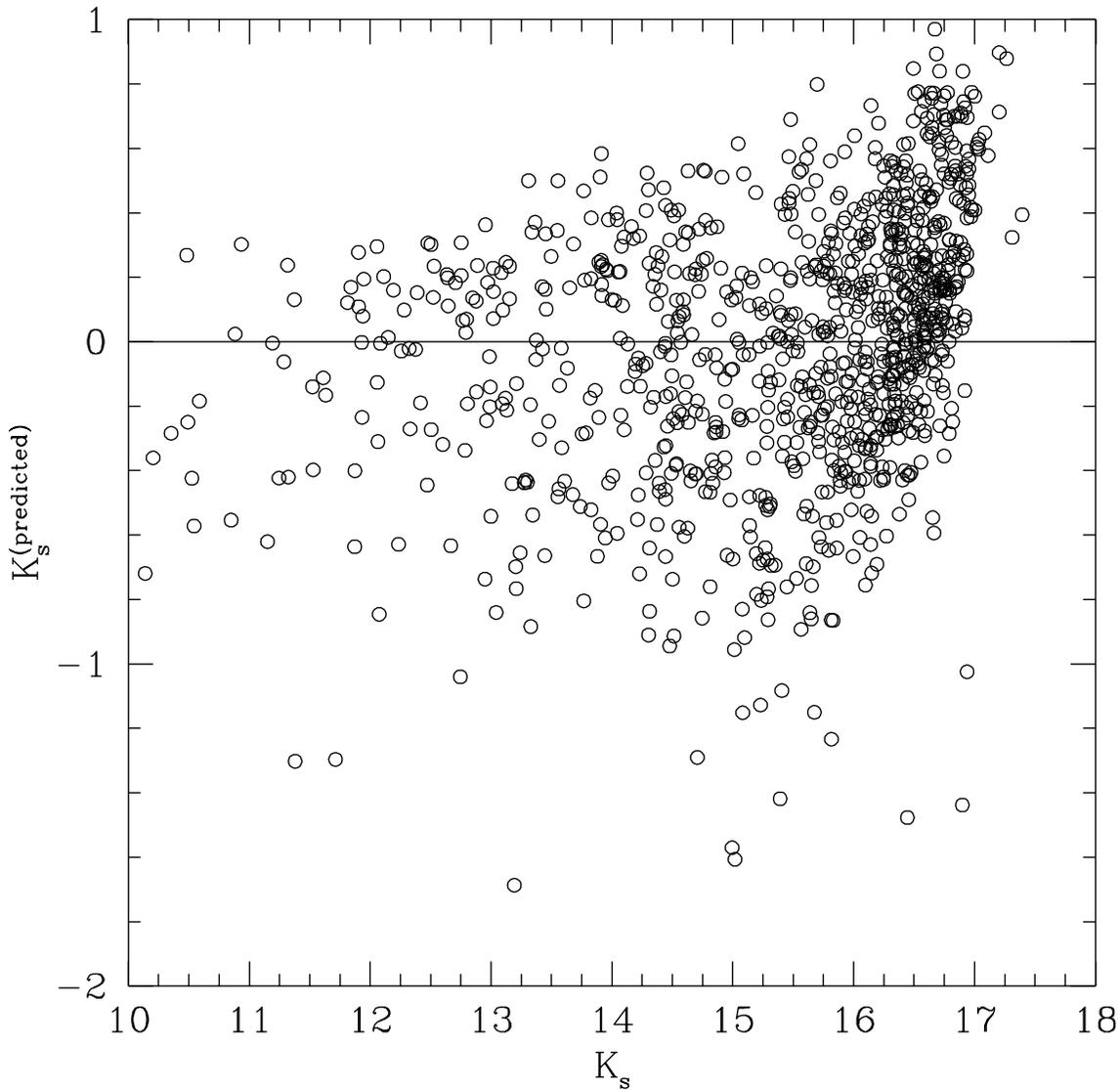}
\caption[kpred.eps]{The correlation between predicted and measure $K_s$
magnitudes. The estimated magnitude is derived by fitting a second
degree polynomial function to the optical $U, J, F, N$ photometry. The rms
dispersion about this relation is 0.4 magnitudes. Using this correlation
the optical and near-infrared data were using a combined angular and magnitude distance metric.}
\end{figure}

\begin{figure}
\plotone{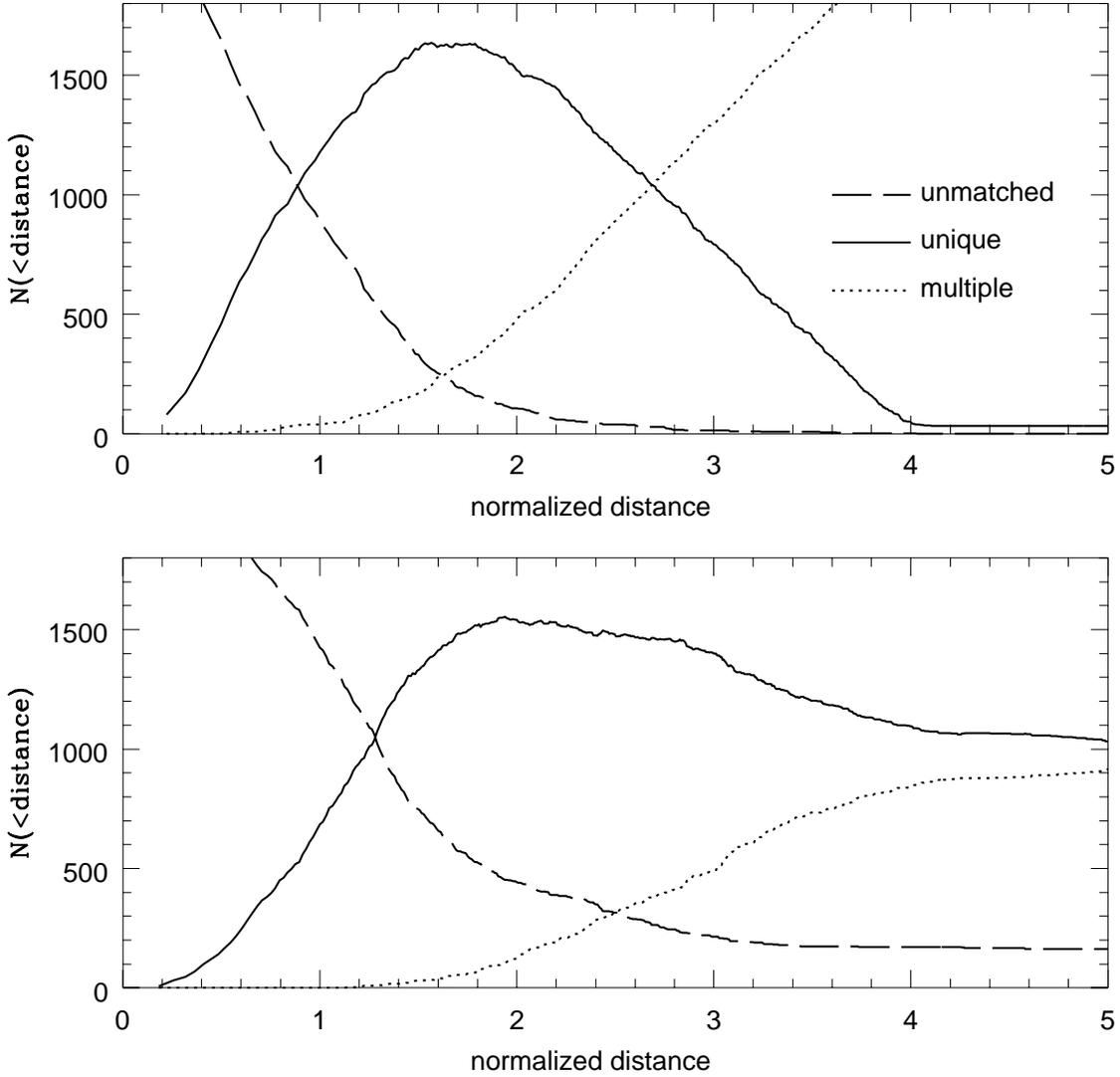}
\caption[pairup.eps]{The effect of using color information to cross-id
catalogs. When one uses angular distance only (top figure), the 
acceptance criteria is very critical. Setting the limit too low will
result in the rejection of valid pairs (dashed curve). Relaxing the
criteria too much results in too many ambiguous pairs (dotted line).
Including the color information eliminates most of the accidental
coincidences (dashed line) and drastically reduces the number of
ambiguous matches. At all distances the ratio of unique to multiple
matchups is larger when we incorporate color information. Further, the
choice of tolerance for the metric is much less important (the solid
line, the number of unique identifications dies off gently). In
both cases we used the dimensionless normalized distance, 
$\rho =\sqrt{{(\alpha_K-\alpha_0)^2\over\sigma_\alpha^2} +
{(\delta_K-\delta_0)^2\over\sigma_\delta^2}}$ and
$\rho =\sqrt{{(\alpha_K-\alpha_0)^2\over\sigma_\alpha^2} +
{(\delta_K-\delta_0)^2\over\sigma_\delta^2} +
{(K-K^\ast)^2\over\sigma_K^2}}$ as defined in section 4.
\label{fig3}}
\end{figure}

\begin{figure}
\plotone{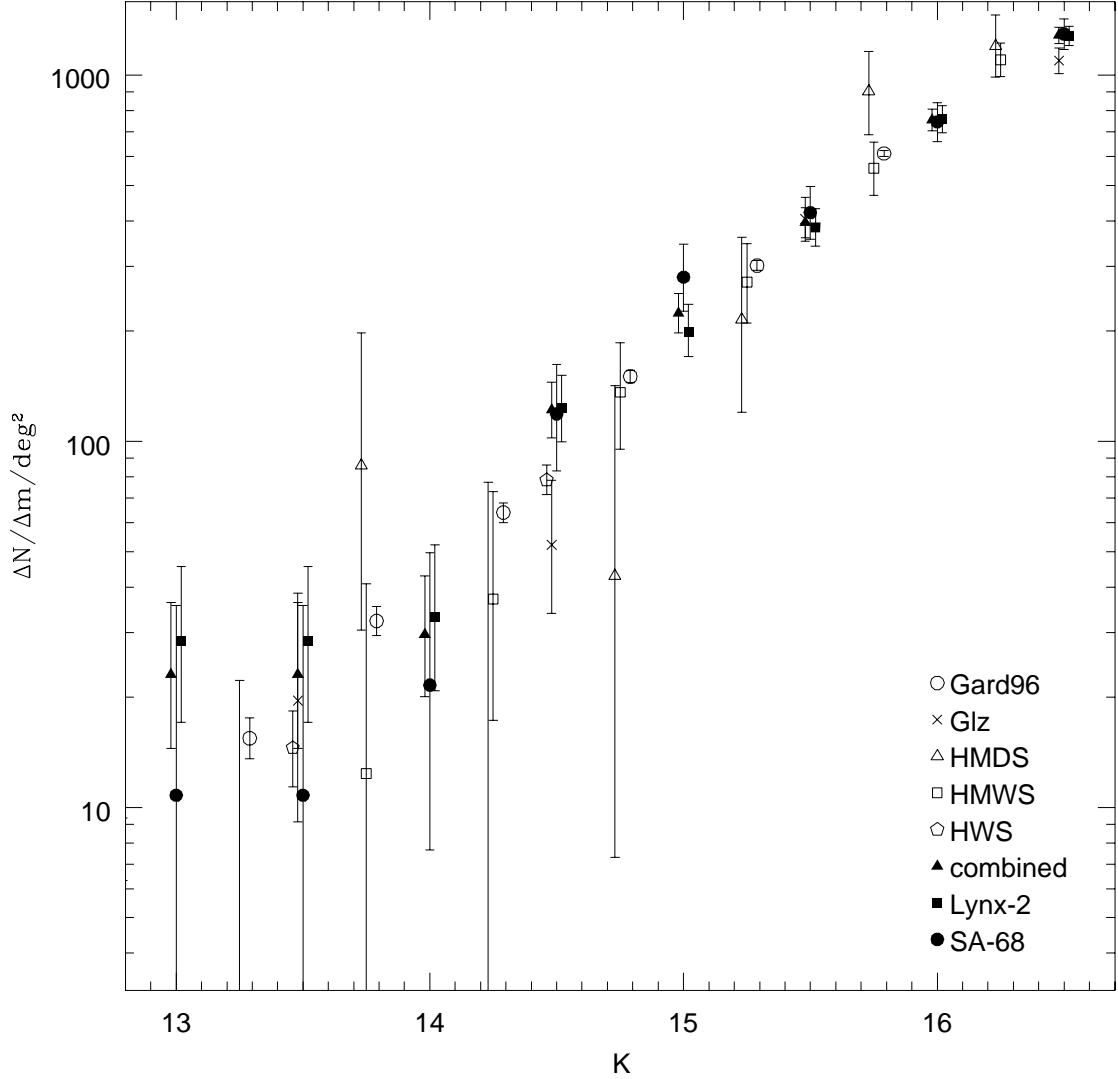}
\caption[ncounts2.eps]{K-band differential number counts in the $K=13 - 16.5$
range. The data are a compilation from \cite{gard93} (HWS, HMWS,
HMDS), \cite{glaz94} (Glz), \cite{gard96} (Gard96), our SA-68 and
Lynx-2 fields, and our two fields combined (combined).  Errors are
$1\sigma$ estimates from the raw counts.
\label{fig4}}
\end{figure}

\begin{figure}
\plotone{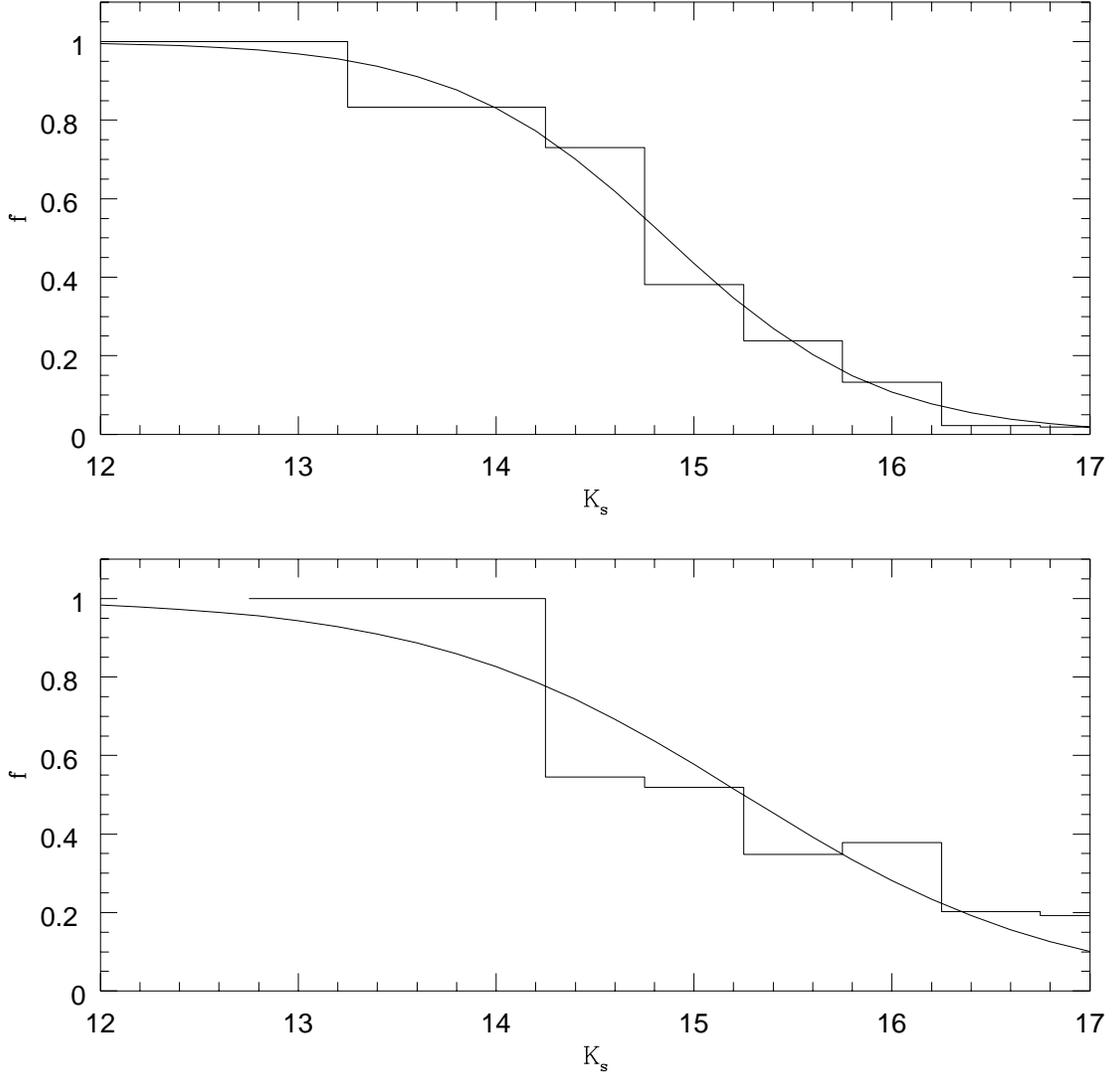}
\caption[comp.eps]{The redshift completeness, $f$, in each of the
fields.  The sample used in the luminosity function is cut at
$K_s=16$ in SA68 and $K_s=15.25$ in Lynx2.  Sandage
functions are fit to the incompleteness with parameters, $m_l=15.25$,
$\Delta m_l=0.80$ in SA68 and $m_l=14.86$, $\Delta m_l=0.54$ in Lynx2
(see text).\label{figcomp}}
\end{figure}

\begin{figure}
\plotone{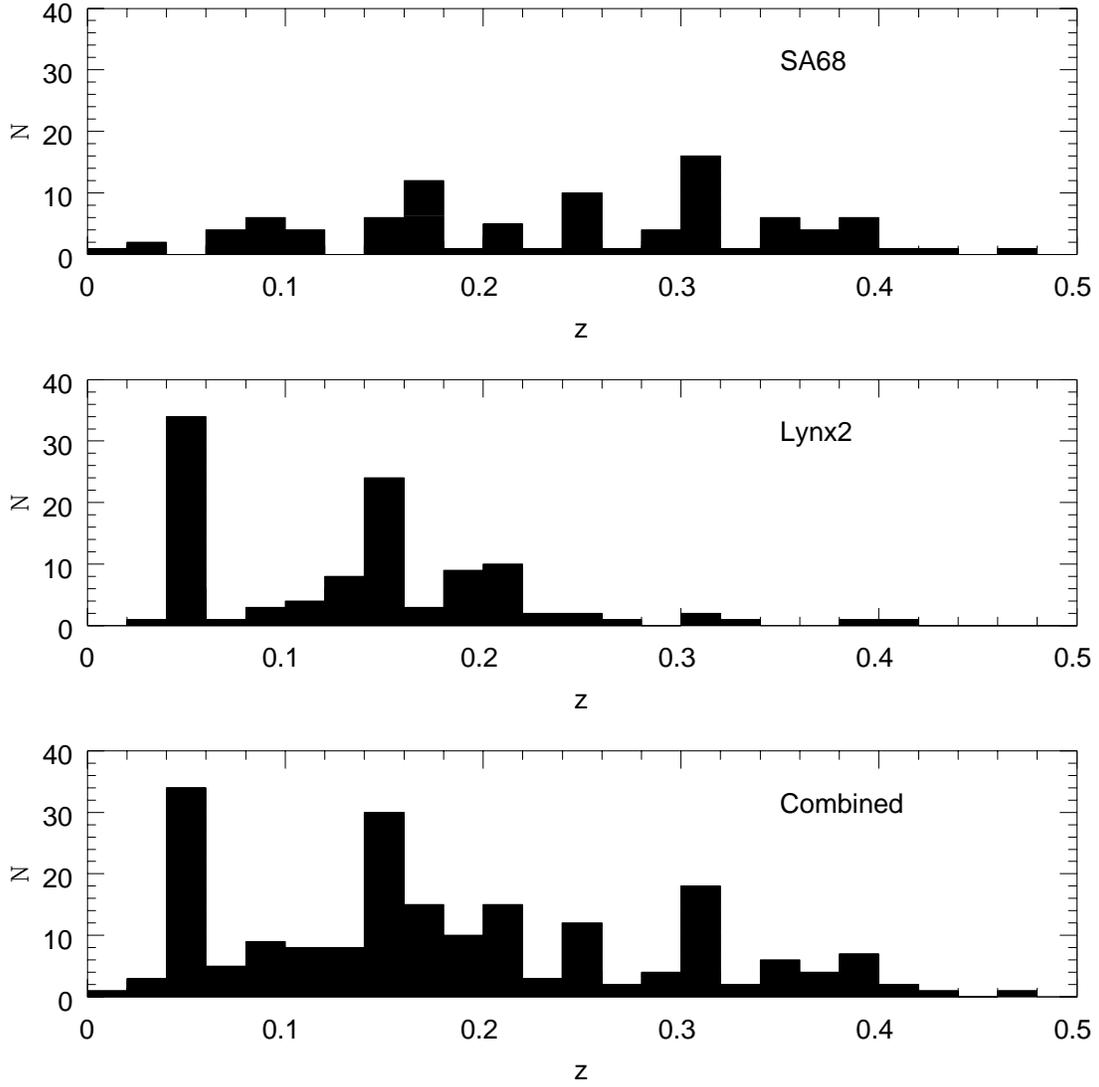}
\caption[zdis.eps]{The redshift distribution of each sample of
galaxies that were used in the luminosity function derivation. Two
points are evident from these distributions. The majority of galaxies
have redshifts $z<0.4$ and the Lynx2 field contains a low redshift
cluster at $z=0.05$. The presence of significant structure in the
redshift distribution requires the use of an estimator of the
luminosity function that is not dependent on the clustering.
\label{figzdis}}
\end{figure}

\begin{figure}
\plotone{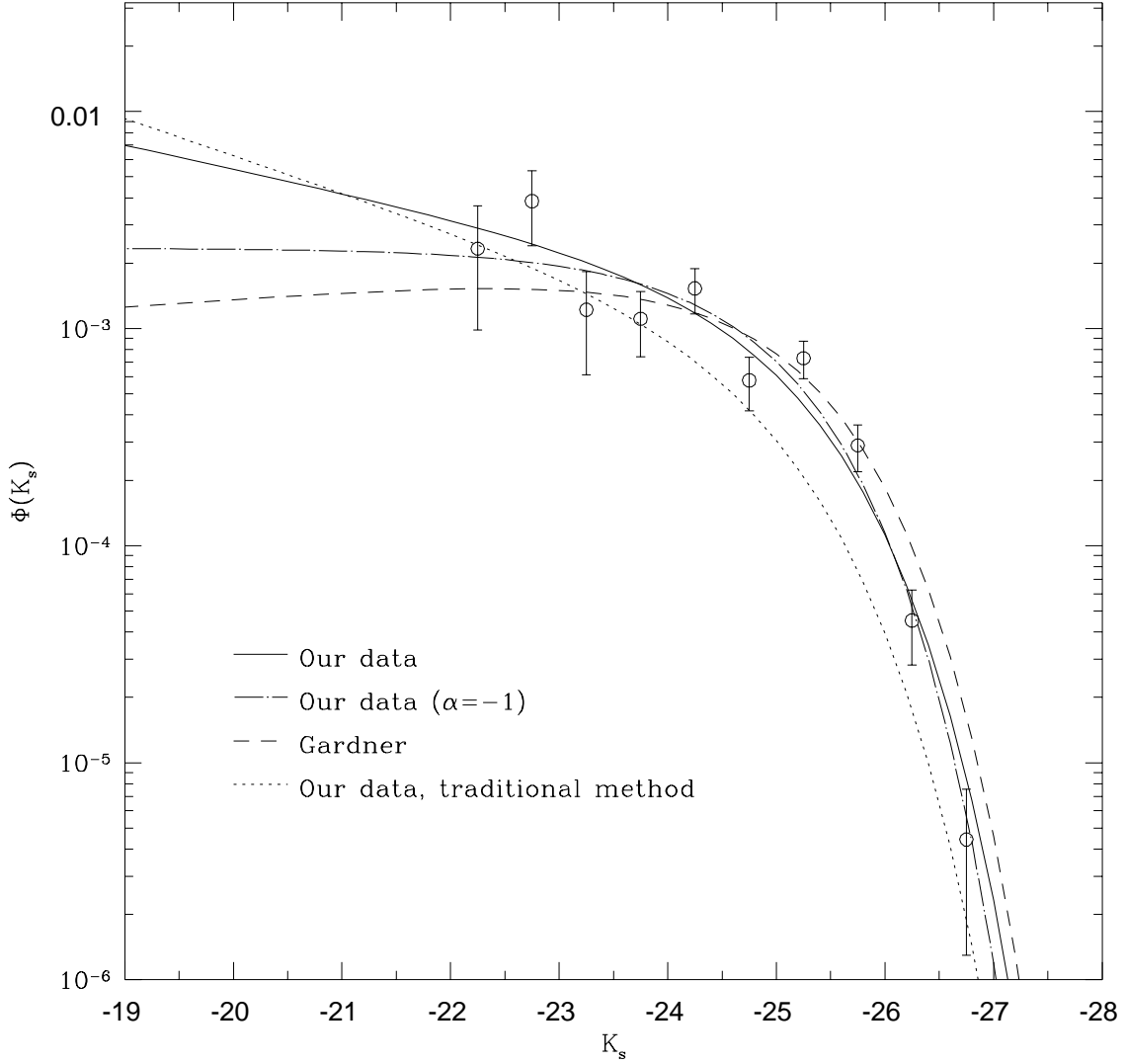}
\caption[lumfun.eps]{Differential luminosity function from our
redshift sample.  Error bars are calculated from the number of
galaxies in each absolute magnitude bin, assuming Poisson
statistics. The lines shown are for the Schechter functions fits (both
with and without setting the parameter $\alpha$ to $-1$) as well as
for previous determinations of the luminosity function.
\label{figlf2}}
\end{figure}

\begin{figure}
\plotone{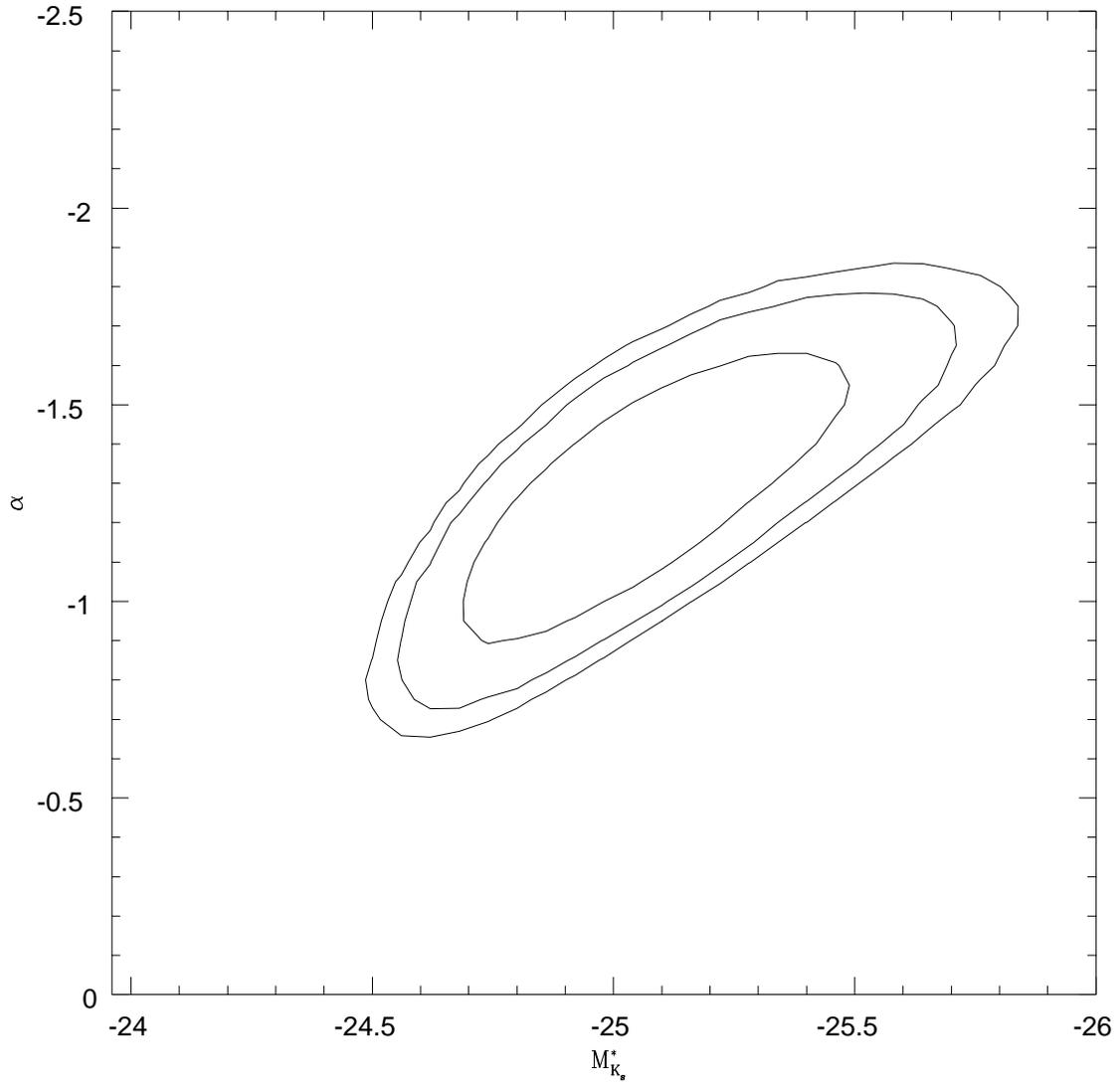}
\caption[contour.eps]{The confidence intervals for the Schechter
parameters. The 68, 90 and 95 percentile confidence interval regions are
shown derived from the maximum likelihood fitting techniques.
\label{figlf1}}
\end{figure}

\end{document}